\documentclass[preprint,aps,pra]{revtex4-1}%revtex4-2 takes too much space and leaves %space between abstract and the authors then use revtex4-1
\usepackage{amsmath,amssymb}
\usepackage[utf8]{inputenc}
\usepackage{amsfonts}
\usepackage{epsfig}
\usepackage{graphicx,xcolor}% Include figure files
\usepackage{bm}% bold math
\usepackage{xcolor}
\usepackage{natbib} %use this? when using bbl in the main text
\usepackage{subcaption}
\usepackage{wrapfig}
\usepackage{lineno}
\usepackage{color}
\usepackage{grffile}
\usepackage{caption}
\usepackage[T1]{fontenc}
\usepackage{url} 
\usepackage[bookmarks=false]{hyperref}
\begin{document}
%\preprint{APS/123-QED}
%\title{Temperature-dependent radiative lifetime measurement: \\ %Molecular Spectroscopy of the Na$_2$ 6$\,^1\Sigma_g^+(9,31)$ State}% 
\title{Temperature-dependent radiative lifetime measurement of the 6$\,^1\Sigma_g^+(9,31)$ state of sodium molecules}%
%%%%%%%%%%%%%%%%%%%%%%%%%%%%%%%%%%%%%%%%%%%%%%%%%%%%
\author{Kshitiz Rai$^1$}
 \author{Shakil Bin Kashem$^1$}\altaffiliation[Also at]{ Washington University in St. Louis} \author{Henry Pierce$^1$}\author{Seth Ashman$^2$}\author{Burcin Bayram$^1$} \email[Contact author:~]{bayramsb@MiamiOH.edu}\affiliation{$^1$Department of Physics, Miami University, 500 E. Spring Street, Oxford, OH}\affiliation{$^2$Department of Engineering, Physics, and Systems, Providence College, Rhode Island, USA}
\begin{abstract}
We report the measurement of radiative lifetimes of the $6^1\Sigma_g^+(v=9,J=31)$ state of gas-phase molecular sodium using a high-resolution double-resonance spectroscopy. Measurements were done using a time-correlated photon counting technique at various pressures and temperatures. Lifetimes were extracted from extrapolations to the zero buffer gas pressure, called Stern-Volmer plot, and the temperature-dependence of the radiative lifetimes were measured  over a temperature range from 593 K to 653 K.  Our result agrees well within the error limits with the theoretical calculations.
\end{abstract}

\maketitle

\section{\label{sec:level1}Introduction}
%\nolinenumbers
Due to the complex internal structure of the molecules, cold and ultracold molecules offer promising platforms for exploring fundamental physics, including symmetry violations and potential extensions to the Standard Model. Their long lifetimes make them strong candidates for the next generation of clock devices~\cite{Benini24}. Thus, lifetime measurements offer insights into molecular stability, coherence properties, and their suitability for precision measurements such as the search for the electric dipole moment of the electron. Measuring the lifetimes of alkali molecules is essential for understanding their energy relaxation dynamics, which directly impacts the efficiency of cooling and trapping techniques. This connection is crucial for advancing cold and ultracold molecular studies, where prolonged coherence times are key to high-precision experiments.

Alkali-metal atoms and their diatomic molecules, known for their simple yet rich internal structures, have long been a focus of theoretical and experimental studies. Their hydrogen-like electronic configurations and convenient transition wavelengths in the visible and near-infrared make them ideal for a wide range of atomic, molecular, and optical physics experiments, as well as for studies in fields such as quantum information and science, astrophysics, plasma physics, and laser physics~\cite{Borgi19,Crossfield15,varshni63}. The investigation of their molecular structure and dynamics, especially through traditional short-range molecular spectroscopy, is important for obtaining key atmospheric parameters and understanding processes like alkali-metal atom-molecule collisions in the ultracold regime. Such collisions have gained importance in controlling chemical reaction rates via Feshbach resonances~\cite{Park2023}.

Accurate knowledge of electronic transition dipole moments (TDMs) as a function of internuclear distance is fundamental for understanding excited-state structures, decay dynamics, and quantum control of molecular dynamics. In the absence of calculated TDMs, experimentally determined lifetimes offer an essential tool for deriving TDMs. We report here radiative lifetime measurements of the gas-phase sodium molecules in the $6^{1}\Sigma_{g}^{+}$(3s + 5s) electronic state. The singlet state potential curves of the sodium molecules were experimentally and theoretically determined~\cite{tiemann1987potential,Tsai1993,magnier94,tsai1994optical1,tsai1994optical2,tsai1994optical3,yurova94,qui2007new,stwalley2016up}, and various techniques have been employed to measure the fluorescence spectrum of sodium molecules, enabling the extraction of molecular constants~\cite{Demtroder69,Demtroder75,Kusch75,Callender76,Babaky89,Chung01,Bayram2023}. Lifetimes of the various intermediate states have been measured~\cite{Baumgartner70,Ducas76,Demtroder76,Mehdizadeh94,Anunciado16,Jayasundara19}.

In this study, we report, for the first time, temperature-dependent radiative lifetime measurements for sodium dimers in the $6^1\Sigma_g^+(v=9,J=31)$ state using a time-resolved double-resonance excitation scheme. Building on earlier measurements that revealed lifetimes shorter than theoretical predictions~\cite{wagle2021}, in this paper we account for collision effects at high temperatures to extract accurate radiative lifetimes, demonstrating an approach to isolate the radiative component in collisional environment. Molecules were excited from thermally populated rovibrational levels of the ground state to the $6^1\Sigma_g^+$ electronic state through the $A^1\Sigma_u^+$ electronic state using two synchronized pulsed lasers. Time-resolved molecular fluorescence was measured using a time-correlated photon counting technique, and the result was compared with theoretical calculations~\cite{Sanli15,Saaranen18,Ashman2024}.

\section{\label{sec:level2}Experiment}
The experimental setup, shown in Fig.~1, used in this study is the same as previously described~\cite{Saaranen18,wagle2021}, and will be briefly outlined here. Sodium metal was placed inside a four-armed stainless-steel heatpipe oven~\cite{vidal69,vidal71}. The center of the heatpipe oven was heated by four-pairs of ceramic electric heaters to at least 300$^o$C to produce sufficient Na$_2$ density ($10^{13}$ molecules/cm$^3$) in the presence of argon as a buffer gas at room temperature to keep the vapor at the center of the heatpipe. In addition, chilled water tubing was wrapped around the ends of the heatpipe oven arms to prevent condensation onto the windows. At about 320$^o$C, estimated atomic and molecular vapor pressures are on the order of 1.39 × 10$^{-2}$ Torr and  1.8 × 10$^{-5}$ Torr, respectively. In this experiment, the primary contributors to perturbing collisions are Na and Ar atoms, as their number density is higher than that of sodium molecules. The vapor pressures of the alkali metals can be determined using the Nesmeyanov formula~\cite{Nesmeyanov63}, and then atomic and molecular number densities can be determined using the ideal gas law equation. The pressure and temperature were measured using a capacitance manometer and a thermocouple probe respectively. 

\begin{figure*}[hbt!]
   \includegraphics[scale = 1.1]{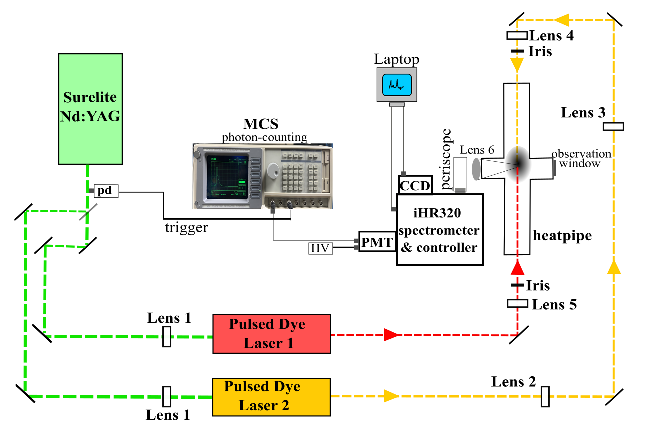}
      \caption{Schematic diagram of the experimental setup. Two pulsed lasers, pumped by a Nd:YAG at 532 nm, are used in counter-propagating configuration for a double resonance excitation. A photomultiplier tube (PMT) and a charge-coupled device (CCD), cooled to -40$^o$C, with 14$\mu$m pixel size, attached to the exit ports of the Horiba imaging spectrometer, are used for signal detection. A multichannel scaler (MCS) is used for time-correlated photon counting in conjunction with the PMT and a photodiode (pd) for triggering the MCS. }
    \label{experiment}
\end{figure*}

Since transition dipole selection rules prohibit transitions between electronic states with the same parity of the electronic wavefunction, we employed double-resonance spectroscopy to study highly excited electronic states with \textit{gerade} symmetry, using the \textit{ungerade} $A^{1}\Sigma_{u}^{+}(10,30)$ state as an intermediate stepping level. The molecules are excited from the thermally populated rovibrational levels of the $X^1\Sigma_{g}^{+}(0,31)$ ground state to the $6^{1}\Sigma_g^{+}(9,31)$ electronic state using counter propagating pulsed laser beams for the time-resolved double-resonance excitation. The excitation scheme is illustrated in Fig.~2. The suitable excitation pathways were identified using the LEVEL8.0 Fortran program code~\cite{LeRoy16}. This program uses experimental or \textit{ab initio} potential energy curves, the transition dipole moment function for the electronic transitions involved, and calculates the important transition quantities such as Einstein coefficients, Franck-Condon factors (FCF),  and energies of the all possible rovibrational transitions. Based on the output of the program code, strong transitions to the $6^{1}\Sigma_{g}^{+}(9,31)$ level were identified and the following excitation paths are selected for the experiment:  $X^1\Sigma_{g}^{+}(0,31)$ $\xrightarrow[\text{FCF = 0.09}]{\text{L1(635.38 nm)}}$ $A^{1}\Sigma_{u}^{+}(10,30)$, and  
 $A^{1}\Sigma_{u}^{+}(10,30)$ $\xrightarrow[\text{FCF = 0.46}]{\text{L2(561.79 nm)}}$ $6^{1}\Sigma_{g}^{+}(9,31)$. 

To reach the $6^{1}\Sigma_{g}^{+}(9,31)$ level, two tunable pulsed dye lasers  in Littman-Metcalf cavity design~\cite{Littman78} were constructed. Both lasers were pumped by a Nd:YAG laser operating at 532 nm at 20 Hz. Laser 1 is tuned to 635.38 nm (15738.62 cm$^{-1}$), which excites molecules from the thermally populated rovibrational level of the $X^1\Sigma_{g}^{+}$ ground state to the $A^{1}\Sigma_{u}^{+}(10,30)$ level. Laser 2 is tuned to 561.79 nm (17800.22 cm$^{-1}$), which excites molecules from the intermediate state to the $6^{1}\Sigma_{g}^{+}(9,31)$ level. Bandwidth of the lasers is less than 6 GHz (1 cm$^{-1}$=30 GHz), the energy separation between vibrational levels is 150 cm$^{-1}$, and rotational levels are 15 cm$^{-1}$, we selectively excite rovibrational levels within the bandwidth of the lasers. The frequency of each laser is measured using a Coherent wavemeter with accuracy of 0.01 cm$^{-1}$.  
%Laser 1 (L$_1$) excites molecules from a thermally populated rovibrational state to %an intermediate $A^{1}\Sigma_{u}^{+}(10,30)$ state. Laser 2 (L$_2$) excites %molecules from the intermediate state to the final $6^{1}\Sigma_{g}^{+}(9,31)$ %state. 

\begin{figure}[hbt!]
\centering
\includegraphics[scale =0.4]{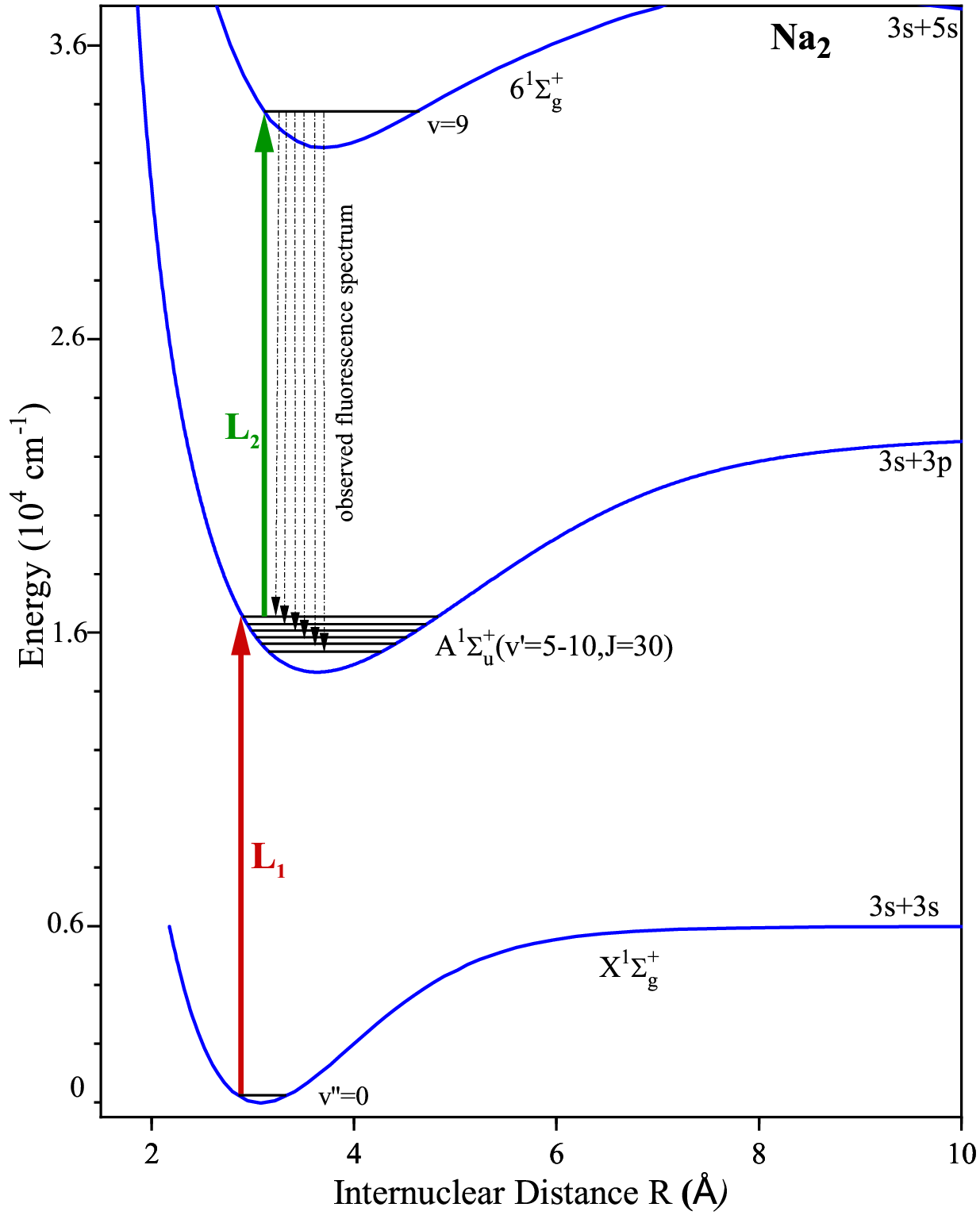}
\caption{Na$_2$ partial potential-energy curves and double-resonance excitation scheme of the Na$_2$ molecule are shown. The Laser 1 (L$_1$) selectively excites molecules from one of the thermally populated rovibrational levels of the $X^1\Sigma_{g}^{+}(0,31)$ ground state to an intermediate rovibrational level of the $A^{1}\Sigma_{u}^{+}(10,30)$  state. Then, the Laser 1 (L$_2$) further excites these molecules to rovibrational levels of the target $6^{1}\Sigma_{g}^{+}(9,31)$ state. Solid vertical lines represent the laser excitations, while the dashed downward lines depict the observed molecular fluorescence from the $6^{1}\Sigma_{g}^{+}(9,31)$ state to the v=5-10, J=30 states of the $A^{1}\Sigma_{u}^{+}$ electronic state. The potential curves are \textit{ab initio} calculations from Ref.~\cite{magnier93}.}
\label{potential}
\end{figure}

\begin{figure}[hbt!]
    \centering
    \includegraphics[scale = 0.4]{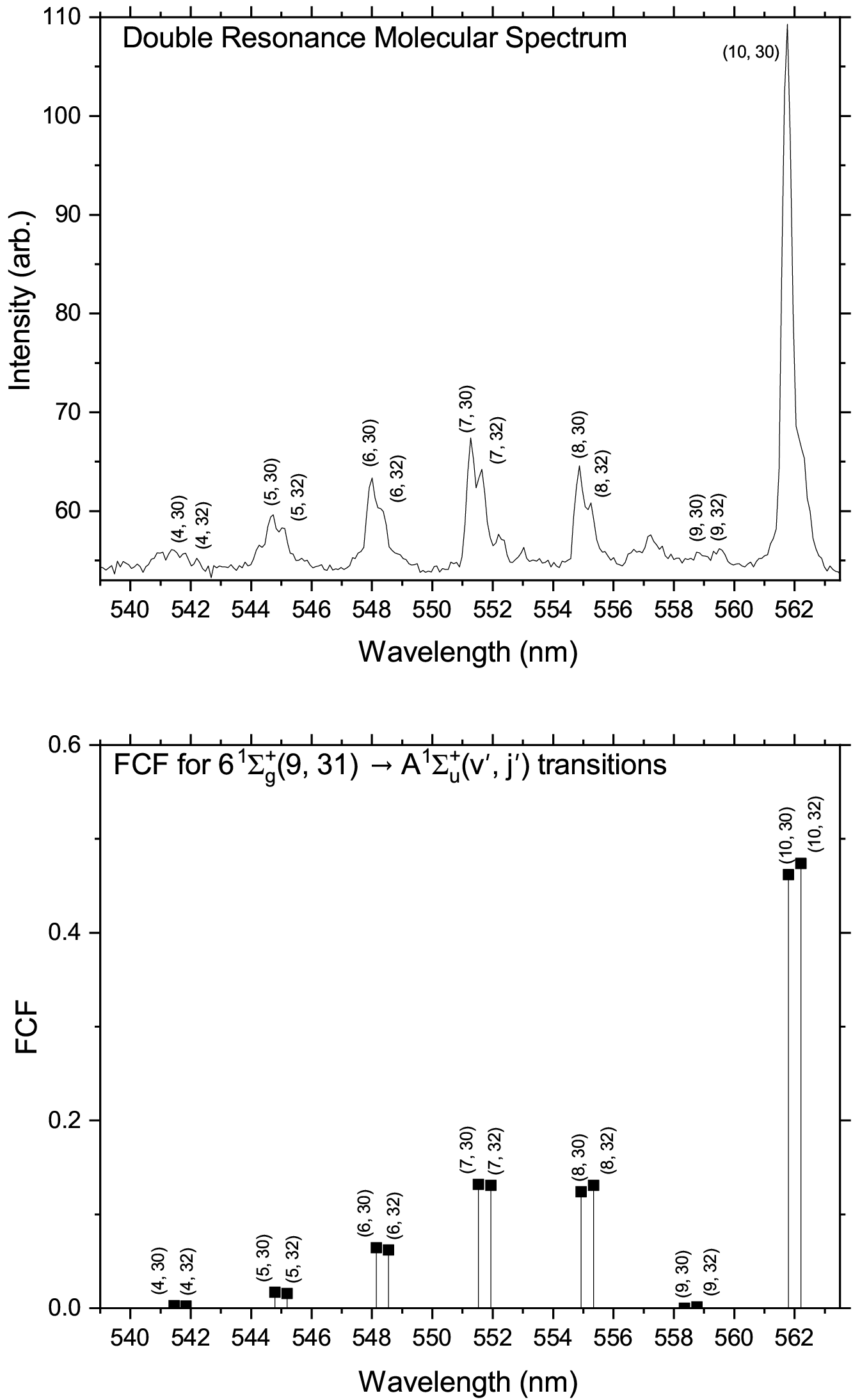} 
    \caption{Comparison of experimentally observed fluorescence spectrum through double-resonance excitation and theoretically calculated FCF spectrum. The spectral peaks consist of $P(32)$ and $R(30)$ branches from the $6^{1}\Sigma_g^{+}(9,31)$ state fluorescence emission. }
    \label{Comparision}
\end{figure}

%\section{Measurements}
Two synchronized pulsed lasers enter the heatpipe oven in counter-propagating direction.  The resolved molecular fluorescence from the $6^{1}\Sigma_g^{+}$ state was collected using Horiba (iHR320) imaging spectrometer which has dual exit ports. One of the exit ports has a cooled charge-coupled device (CCD) which is used to collect laser induced molecular spectrum and the other side of the exit port has a photomultiplier tube (PMT, R928P Hamamatsu Photonics). The CCD is cooled to -40$^o$C and has 14$\mu$m x 14$\mu$m pixel size with 2048 x 70 array dimension. The PMT is connected to the Multichannel Scaler (SR430 MCS) which is triggered by a laser pulse using a photodiode (SM05PD1A Thorlabs) and thus MCS scans count the PMT pulses. The trigger threshold was set to 100 mV 
by the rising edge of the photo-diode pulse. The PMT has a typical rise time of 2.2 ns, which is suitable given that the minimum bin width of the MCS is 5 ns. With the discriminator threshold set at 100 mV, the MCS can reliably time PMT pulses within the 5 ns bin width. The FWHM resolution of the spectrometer, with the 1200 grooves/mm holographic grating, at the PMT side exit port with slit width at 90x10$^{-6}$ m is 19.33 cm$^{-1}$ and at the CCD side is 3 cm$^{-1}$. Figure~\ref{Comparision} shows experimentally observed double-resonance spectrum and theoretically calculated FCF spectrum. The spectral peaks consist of $R(30)$, $P(32)$ branches from the $6^{1}\Sigma_g^{+}(9,31)$ state to the $A^{1}\Sigma_u^{+}(v'=5-10,J')$ were observed. To identify and confirm the assignment of the spectra, we used LEVEL program output and compared the measured fluorescence spectrum with FCF spectrum. The thermal energy $k_BT$ of the heatpipe oven is at about 454 cm$^{-1}$ which is a few vibrational separations, thus it is possible for the collisional-induced transitions with Na, Na$_2$, and Ar to populate nearby rovibrational energy levels which then emit fluorescence to contribute additional spectral features to the spectrum. 

%%%%%%%%%%%%%%%%%%%%%%%%%%%%%%%%%%%%%%%%%%%%%%
Since sodium nuclei ($I$=3/2) are fermions, they obey Pauli-Exclusion principle which requires that the total wave function of the molecule must be antisymmetric. There are a total of (2$I$+1)$^2$=16 configurations to combine the spins of the two $\,^{23}$Na sodium nuclei. Thus, the possible values of the total nuclear spin configurations ($I$=3, 2, 1, 0) are ten antisymmetric nuclear-spin wave functions, which form ortho-states ($I$=1,3 and nuclear spin weight g=10), and six symmetric nuclear-spin wave functions, which form para-states ($I$=2,0 and nuclear spin weight g=6). Therefore, for the total wave function to be antisymmetric, the antisymmetric rotational levels (odd-$J$) of the $X\,^{1}\Sigma_g^{+}$  electronic state must combine with the ortho nuclear spin configurations while symmetric rotational levels (even-$J$) of the $X\,^{1}\Sigma_g^{+}$  state will combine with the para nuclear spin states~\cite{Bernath16,atkins2013,Herzberg50,Duck98}. Since the total symmetry of the molecular wave function remains unchanged, the molecules formed in the heat-pipe oven has a mixture of odd-$J$ and even-$J$ configurations, and thus all rotational levels of the ground electronic state are populated. Hence, optical transition out of any rotational level can be selected. Thermal distribution of the rotational levels of the ground electronic state is a curve which goes through a maximum $J$ first as temperature increases ($J_{max}=0.5896 \sqrt{T/B_e}-0.5$, B$_e$=0.1547 cm$^{-1}$)~\cite{Herzberg50,Radzig85,Bayram2023}. Since some of the highly populated rotational levels of the $X\,^{1}\Sigma_{g}^{+}$ electronic state at the operating temperature of the heatpipe are in the range $J$=26-35, the transition out of the $X\,^{1}\Sigma_{g}^{+}(9,31)$ was selected for this experiment.  

%%%%%%%%%%%%%%%%%%%%%%%%%%%%%%%%%%%%%%%%%%%%
For the lifetime measurement, the fluorescence peak corresponding to R(30), the transition $6^{1}\Sigma_g^{+}(9,31) \xrightarrow{} A^{1}\Sigma_{u}^{+}(8,30)$, chosen for its strong intensity, was directed to the PMT-side exit slit of the spectrometer. The signal from the PMT was then sent to the MCS for time-correlated photon counting. The MCS tracks and records PMT input pulses across time bins at a 5 ns resolution, triggered by a 100 mV laser pulse. For each trigger pulse, the MCS records 20,000 data points per scan and displays on the screen on-real time. Then, a decay histogram is created by compiling these records. From the decay histogram of the MCS, we determine the lifetime of the excited molecules by fitting the data with an exponentially modified Gaussian peak function using Gaussmod fit in Origin 2023 software.

To determine collision-free lifetime, the molecular fluorescence from the $6^{1}\Sigma_g^{+}(9,31)$ state was recorded at various pressures with a set temperature ranging from $320^{\circ}$C to $380^{\circ}$C, as shown in Fig.~4. The horizontal error bars represent the error in the pressure measurement while the vertical errors are from the Gausmod fitting.  The radiative and non-radiative lifetimes are related by the Stern-Volmer relationship~\cite{Stern19,baumgartner84}. The inverse of the effective lifetimes against argon pressures were plotted using Stern-Volmer relationship and the effect of the collisions with argon is eliminated with the Stern-Volmer extrapolation at each temperature setting. We found the argon-collision free lifetimes at $320^{\circ}$C to be 27.2 $\pm$ 0.6 ns, at  $340^{\circ}$C to be 26.3 $\pm$ 1.0 ns, at  $360^{\circ}$C to be 25.4 $\pm$ 0.7 ns, and at  $380^{\circ}$C to be 23.9 $\pm$ 0.6 ns.
%\noindent The effected decay rate, $k_{e}$, from the $6^{1}\Sigma_g^{+}(9,31)$ state can be written as $k_{e}=k_T + (k_P)$, where $k_P$ is due to the collisions with Ar atoms and $k_T$ is due to collisions with Na-Na$_2$ collisons, respectively. Stern-Volmer plots yields lifetimes at the argon collision-free regime. 
%%%%%%%%%%%%%%%%%%%%%%%%%%%%%%%%%%%%%%%%%%%%
\begin{figure*}[ht!]
\begin{subfigure}[t]{.5\textwidth}
  \centering
  \includegraphics[width=1\linewidth]{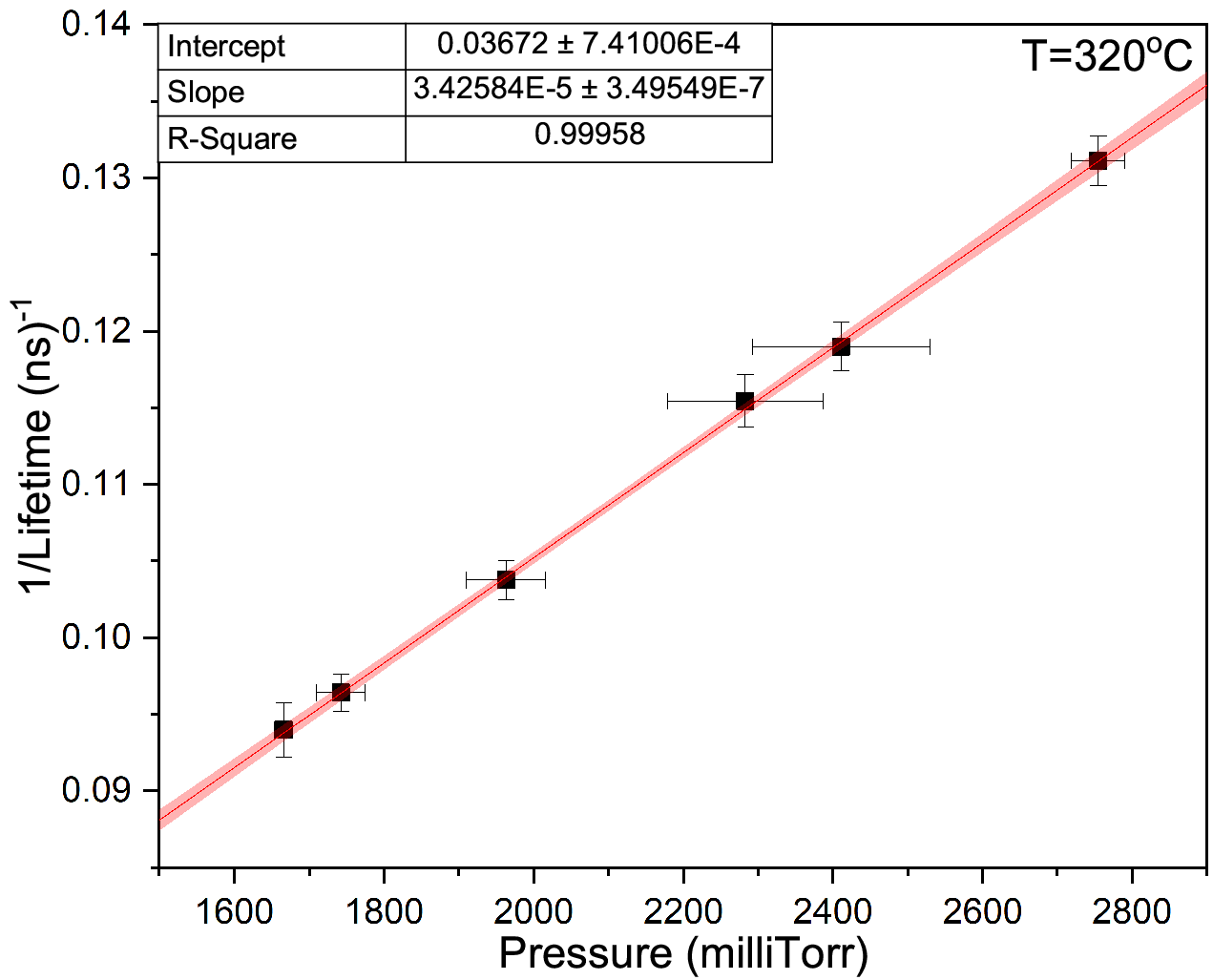}
  \caption{Lifetime measurement at temperature of 320$^\circ$C was determined to be $27.2 \pm 0.6$ ns.}
  \label{320}
\end{subfigure}\hfill
\begin{subfigure}[t]{.5\textwidth}
    \centering
  \includegraphics[width=1\linewidth]{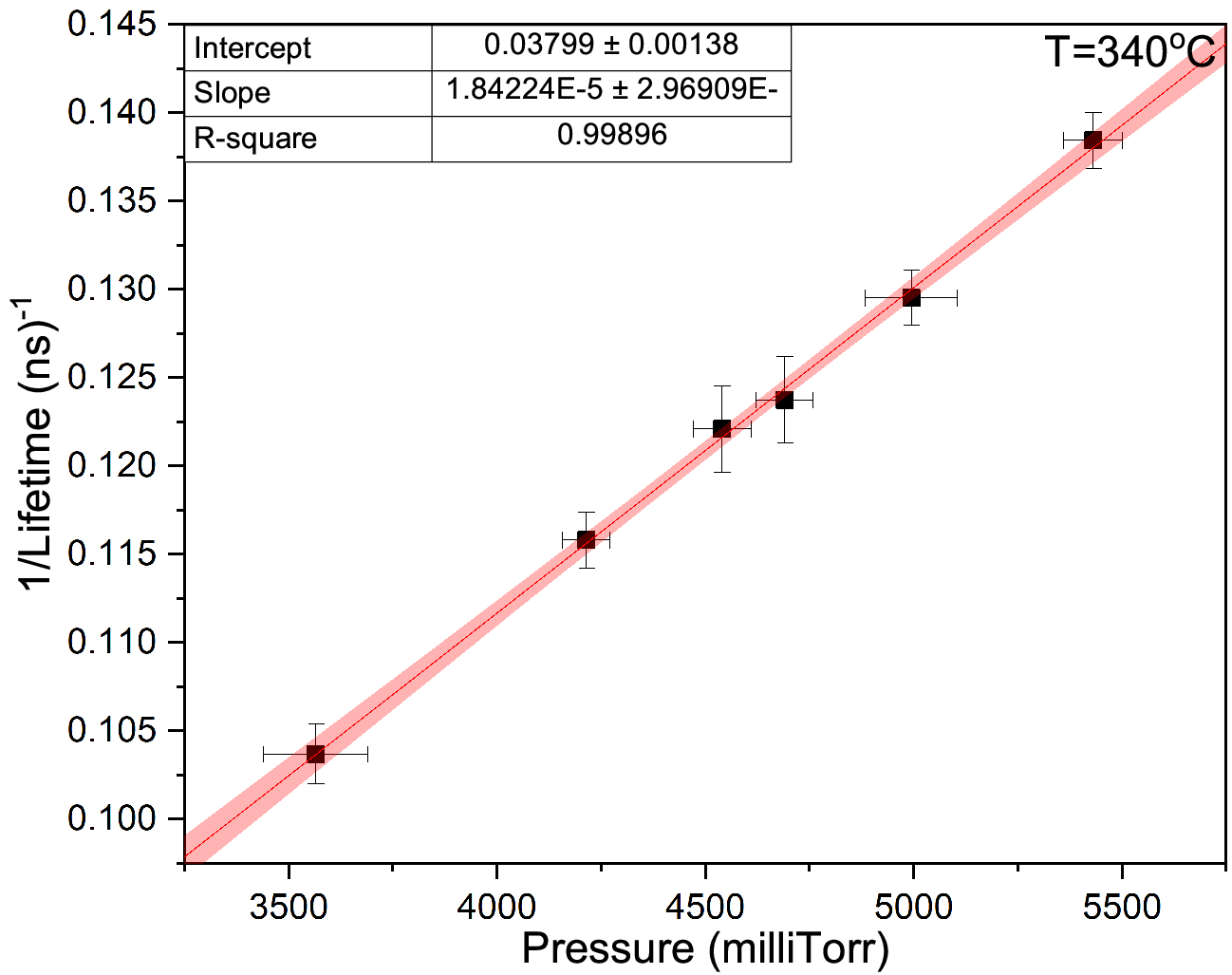}
  \caption{Lifetime measurement at temperature of 340$^\circ$C was determined to be $26.3 \pm 1.0$ ns.}
  \label{340}
\end{subfigure}
\medskip % create some *vertical* separation between the graph
%%%%%%%%%%%%%%%%%%%%%%%%%%%%%%%%%%%%%%%%%%%%%%%%%%%%
\begin{subfigure}{.5\textwidth}
    \centering
  \includegraphics[width=1\linewidth]{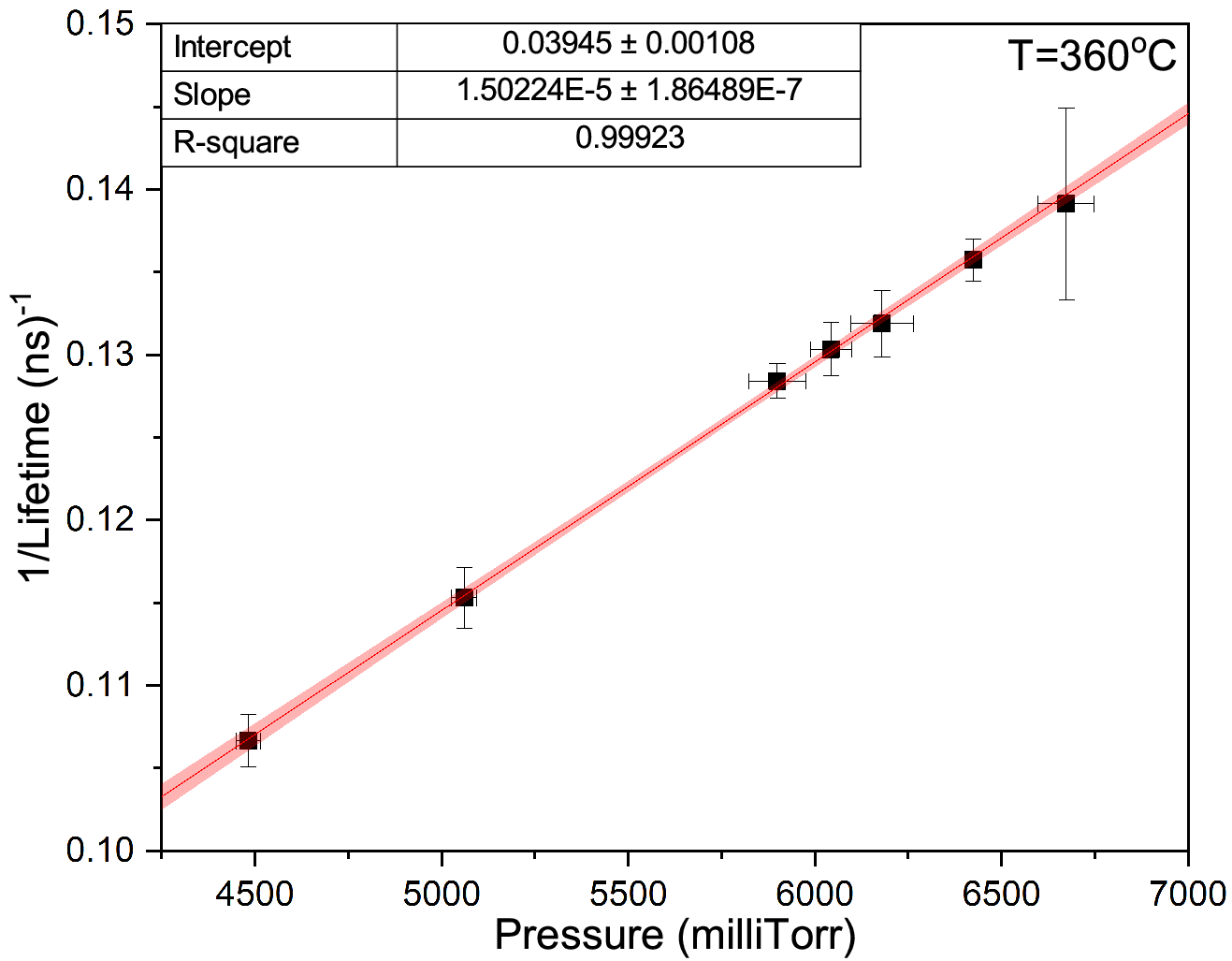}
  \caption{Lifetime measurement at temperature of 360$^\circ$C was determined to be $25.4 \pm 0.7$ ns.}
  \label{360}
\end{subfigure}\hfill
\begin{subfigure}{.5\textwidth}
    \centering
  \includegraphics[width=1\linewidth]{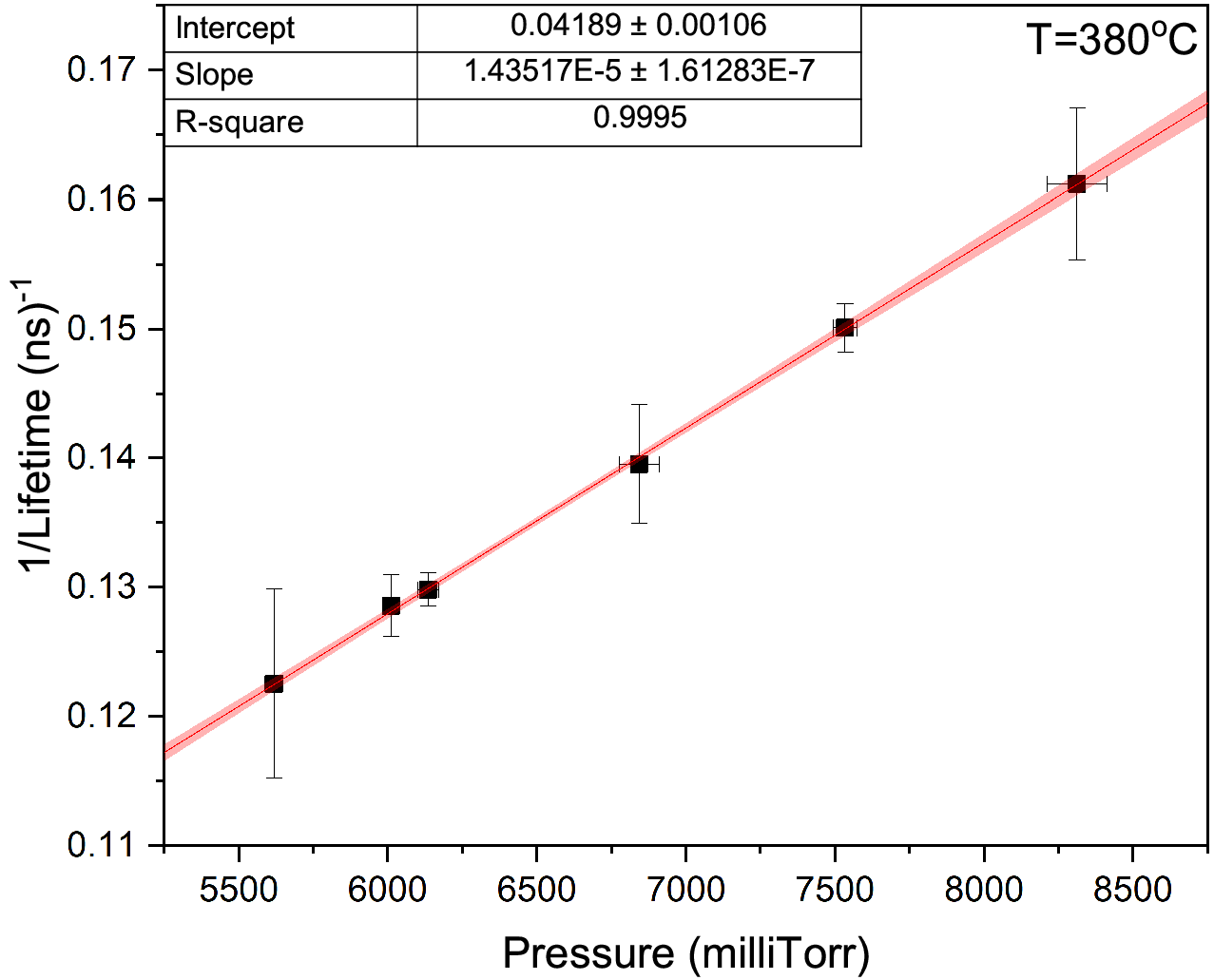}
  \caption{Lifetime measurement at temperature of 380$^\circ$C was determined to be $23.9 \pm 0.6$ ns.}
  \label{380}
\end{subfigure}
\caption{Stern-Volmer plots at four different temperature settings are shown. Argon-collision free lifetimes for the $6^1\Sigma_{g}^{+}(9,31)$ state were determined by extrapolating a linear fit to the data. The red shaded part shows the 95\% confidence band. }
\label{Ar-free lifetime}
\end{figure*}
%%%%%%%%%%%%%%%%%%%%%%%%%%%%%%%%%%%%%%%%%%%%%%%%%%%%%%%%%%%%%%
\section{Results}

\begin{figure}[h]%[ht!]
    \centering
    \includegraphics[scale = 0.37]{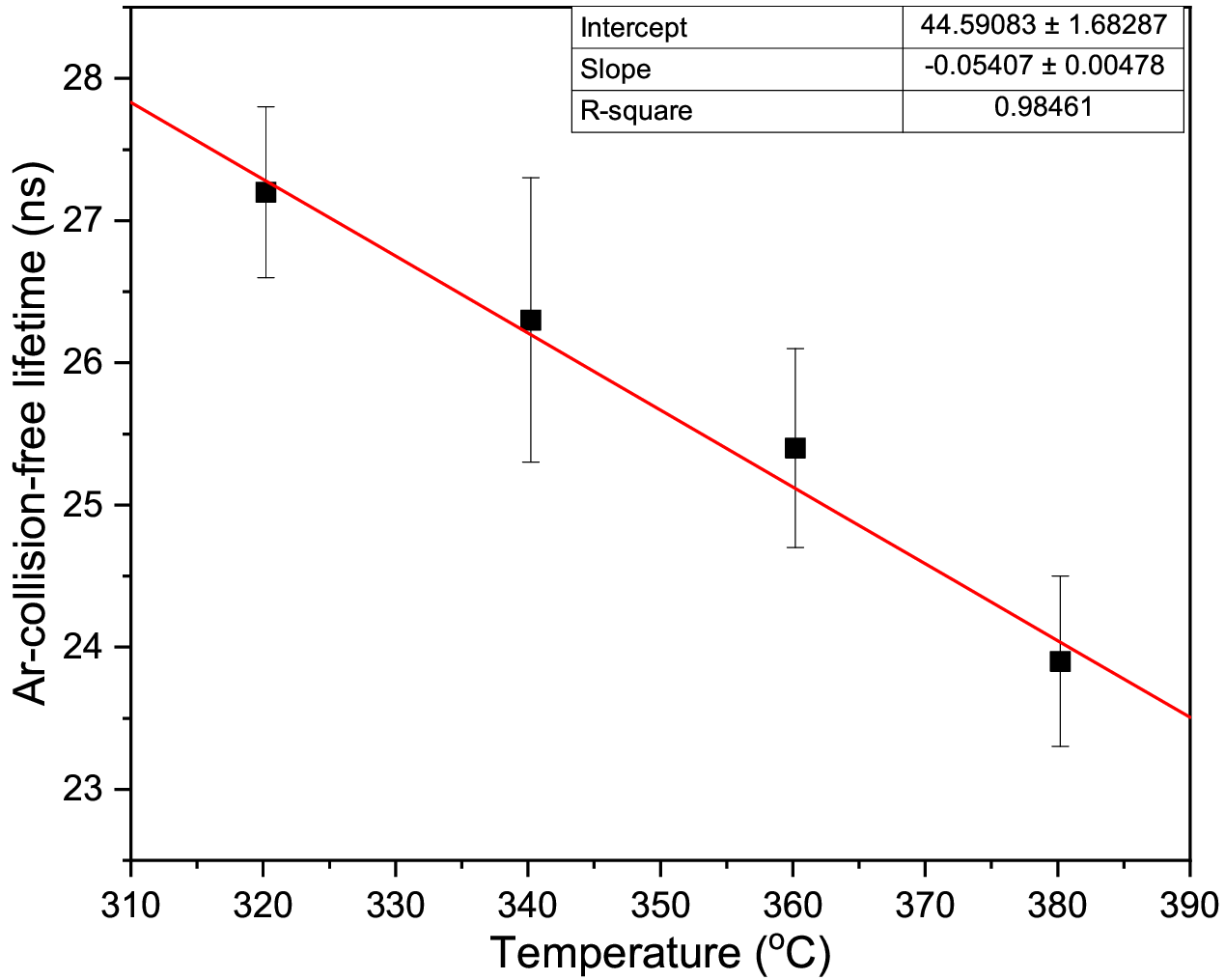}
    \caption{The radiative lifetime of the $6^{1}\Sigma_g^{+}(9,31)$  state was determined by using a linear fit to the data. The result yields a value of 43.5 $\pm$ 1.7 at 20$^o$C (room temperature). }
    \label{Radiative lifetime}
\end{figure}

From the Stern-Volmer plots in Fig.~4, we observed a temperature dependency in the measured lifetimes. Due to the high operating temperatures of the heat-pipe oven, the system may be in a non-equilibrium state, leading to quenching of the alkali atomic vapor density and consequently affecting the lifetimes. The sodium atom number density increases from 1.3 × 10$^7$ at 593.15 K (320°C) to 6.3 × 10$^9$ at 653.15 K (380°C). In comparison, the sodium molecular density at 380°C is 2.7 × 10$^6$, making the atomic density approximately on the order of three times higher. Therefore, the observed effect on the lifetimes is more likely attributed to Na$_2^*$-Na collisions. To account for the influence of these collisions, we plotted the argon-collision-free lifetimes against their corresponding temperatures, as shown in Fig.~\ref{Radiative lifetime}. As expected, lifetimes decrease with increasing temperature due to the Na$_2^*$-Na collisions.   

The radiative lifetime of the $6^{1}\Sigma_g^{+}(9,31)$  state was determined by extrapolating a linear fit to the data. The result yielded a value of 43.5$\pm$1.7 ns at 20$^o$C with a correlation coefficient of 0.98. The experimental datasets generated and analyzed during this study are available in the Harvard Dataverse repository~\cite{BayramDataverse1}. The radiative lifetime is simply the inverse of the total decay rate (Einstein $A$ coefficients) of the excited level to all allowed rovibrational energy levels. For lifetime calculations, we first used LEVEL8.0 Fortran program code~\cite{LeRoy16} for the bound-bound transition to solve the Schr{\"o}dinger equation numerically. Then, we used BCONT program~\cite{LeRoy2004bcont} for the bound-free transitions by entering the transition dipole moment functions for the two interested molecular potential energy curves. The spontaneous decay from the $6^{1}\Sigma_g^{+}(9,31)$ state to all seven allowed electronic potential transitions are $A\,^1\Sigma_u^+$~\cite{Tiemann96}, $B\,^1\Pi_u^+$~\cite{Camacho05}, $2^1\Pi_u^+$~\cite{Grochola05}, $3\,^1\Pi_u^+$~\cite{Grochola06}, $2^1\Sigma_u$~\cite{Pashov00-1}, $3^1\Sigma_u$, and $4^1\Sigma_u$~\cite{Grochola04} states. The details of similar calculations were explained in earlier work~\cite{Saaranen18}. Since the summation over all possible rovibrational decay rates yields the Einstein $A$ coefficient, the expression for the radiative lifetime can be written as $\tau_k$=$A_k^{-1}$=$(\Sigma A_{ki})^{-1}$, where $k$ denotes the  $6^{1}\Sigma_g^{+}(9,31)$ state. The result of the calculated radiative lifetime for the $6^{1}\Sigma_g^{+}(9, 31)$ state is found to be 43.334 ns~\cite{Ashman2024}. When bound-free transition is not included, the lifetime value increases to about 130 ns~\cite{Sanli15}. For the $6^{1}\Sigma_g^{+}(9,31)$ state, the agreement between the experimentally obtained value 43.5$\pm$1.7 ns and the theoretical value 43.334 ns appears to be in excellent agreement within the experimental error limits. 
\section{Conclusion}
This study reports a high-resolution experimental study of radiative lifetimes in the highly excited $6^{1}\Sigma_g^{+}(9, 31)$ state of sodium dimers using time-correlated photon counting techniques. We precisely measured the argon pressure- and temperature-dependent lifetimes. Lifetimes were extracted through Stern-Volmer extrapolations to zero buffer gas pressure, covering a temperature range of 593 to 653 K. By analyzing the Stern-Volmer plots, we isolate the radiative decay rate from collisional contributions. This approach allows for a more precise determination of radiative properties, which are fundamental for understanding molecular energy relaxation dynamics in gas-phase alkali dimers. Our experimental findings are in excellent agreement with theoretical calculations, which include both bound-bound and bound-free transitions across all allowed electronic states. 

%Thus, the results emphasize the importance of bound-free transitions in the lifetime calculations. 

%We measured the argon pressure- and temperature-dependent lifetimes and extracted the radiative lifetime of the Na$_2$ $6^1\Sigma_g^+(9,31)$ state using a time-resolved double-resonance spectroscopy. In this study, we incorporate collision effects at high temperatures to accurately determine radiative lifetimes, presenting an improved approach for isolating the radiative decay rate from collisional contributions.  Theoretical calculations are done including both bound-bound and bound-free transitions from the target state to all allowed electronic potential transitions. Our experimental results are in good agreement with the theoretical calculations of Ref.~[30, 31]. 

\section*{Acknowledgments}
This work was supported by National Science Foundation Grant Nos. NSF PHY-2309340 and PHY-1607601. 

%\bibliography{main.bib}


%%%%%%%%%%%%%%%%%%%%%%%%%%%%%%%%%%%%%%%%%%%%%%%%%%%%%%%%%%%%%%%%%%%%%%%%%%%%

\end{document}